\DeclareUnicodeCharacter{2212}{\textminus}
\documentclass[
    reprint, twocolumn,
    aps,prb,10pt,amsmath,amssymb,
    longbibliography,superscriptaddress,
    showpacs,preprintnumbers
]{revtex4-2}

\newcommand{\mkq}[0]{{m\mathbf{k}+\mathbf{q}}}
\newcommand{\nk}[0]{{n\mathbf{k}}}

\newcommand{\qnu}[0]{{\mathbf{q}\nu}}
\newcommand{\nnnl}{\nonumber \\}

\usepackage[colorlinks,bookmarks=false,citecolor=blue,linkcolor=blue,urlcolor=blue]{hyperref}
\usepackage{epsfig}
\usepackage{float}
\usepackage{mathtools}
\usepackage[section]{placeins}
\usepackage{cancel}
\usepackage{orcidlink}
\usepackage{xcolor}

\allowdisplaybreaks

\begin{document}

\title{Non-adiabatic phonon self-energy due to electrons with finite linewidths}

\author{Cheol-Hwan Park\,\orcidlink{0000-0003-1584-6896}}
\email{cheolhwan@snu.ac.kr}
\affiliation{Department of Physics and Astronomy, Seoul National University, Seoul 08826, Korea,}
\affiliation{Center for Theoretical Physics, Seoul National University, Seoul 08826, Korea}

\date{\today}

\begin{abstract}
We develop the theory of the non-adiabatic phonon self-energy arising from coupling to electrons with finite linewidths using the spectral representation of Green's functions. Our formalism naturally includes the contribution from the intra-band electronic transitions (as well as the inter-band ones) at all electron wavevectors to the phonon linewidths, which is forbidden for zone-center optical phonons if infinitesimal electron linewidths are used. As a proof of principle, we use the theory to calculate the linewidth of the double-degenerate, zone-center optical $E_{\rm 2g}$ phonons of graphene as a function of the chemical potential. The calculated phonon linewidths obtained with finite electron linewidths differ significantly from those obtained with infinitesimal electron linewidths even at low temperatures. Intra-band electronic transitions play an important role in making this difference. Moreover, only the results accounting for the finite electron linewidths are in quantitative agreement with the available experimental data. The presented formalism is suitable for first-principles calculations.
\end{abstract}

\maketitle

\section{Introduction}

The phonon energy is conventionally obtained from density-functional perturbation theory calculations~\cite{2001Baroni_RMP_DFPT} within the adiabatic, Born-Oppenheimer approximation. Many-body interaction effects beyond the adiabatic approximation can be incorporated as energy-dependent, complex self-energy arising from electron-phonon interactions. This self-energy is referred to as the non-adiabatic phonon self-energy~\cite{2017Giustino_RMP_ElPh}.

Non-adiabatic phonon self-energy plays an important role in materials in which the energy difference between occupied and empty electronic states is comparable to phonon energies. Some of the early examples of first-principles calculations on the imaginary part of the non-adiabatic phonon self-energy arising from electron-phonon interactions are \citet{PhysRevB.19.3708, PhysRevB.57.11276, PhysRevLett.90.095506, PhysRevB.73.155426}.

In previous first-principles studies, the imaginary part of the non-adiabatic (NA) phonon self-energy was calculated as suggested by ~\citet{1972Allen_PRB_phonon_linewidth}:
\begin{widetext}
\begin{equation}
\label{eq:Pi_omega_non_adiabatic_Allen}
{\rm Im\,} \Pi^{\rm NA}_{\qnu}(\omega_\qnu)=-\pi\,\sum_{mn}\,\int \frac{d{\bf k}}{\Omega_{\rm BZ}} 
|g_{mn\nu}({\bf k},{\bf q})|^2\,\left[f(\varepsilon_\nk)-f(\varepsilon_\mkq)\right]\,
\delta (\varepsilon_\nk+\omega_\qnu-\varepsilon_\mkq)\,,
\end{equation}
\end{widetext}
where $f(\varepsilon)=1/\left[e^{(\varepsilon-\mu)/k_{\rm B}T}+1\right]$ is the Fermi-Dirac occupation factor at temperature $T$ and chemical potential $\mu$, and $\Omega_{\rm BZ}$ is the volume of the Brillouin zone.
The electron-phonon coupling constant is given by
\begin{equation}
\label{eq:g}
g_{mn\nu}({\bf k},{\bf q}) = \left<\mkq\left|\Delta V_\qnu\right|\nk\right>\,,
\end{equation}
where $\left|\nk\right>$ indicates the electronic state with wavevector ${\bf k}$, band index $m$ and energy $\varepsilon_\nk$, and $\Delta V_\qnu$ is the change in self-consistent potential due to a phonon with branch index $\nu$, wavevector ${\bf q}$, and energy $\omega_\qnu$.

The energy conservation condition
\begin{equation}
\label{eq:energy_conservation}
\varepsilon_\mkq=\varepsilon_\nk+\omega_\qnu
\end{equation}
was imposed in calculating the imaginary part of the non-adiabatic phonon self-energy because of the energy delta function in Eq.~\eqref{eq:Pi_omega_non_adiabatic_Allen}.
Therefore, this condition is unnecessary if we consider finite electron linewidths. It was shown that such finite electron linewidths play an important role in the non-adiabatic phonon self-energy~\cite{PhysRevB.9.4733,PhysRevB.45.9865,maximov1996,PhysRevB.73.140505,PhysRevLett.100.226401,PhysRevB.96.054515,ferrante2018, PhysRevB.98.041112,Novko2020,Novko2025}.
Among these studies, \citet{PhysRevB.98.041112,Novko2020} and \citet{Novko2025} considered the effect of electron linewidth for the intra-band scattering processes by including higher-order electron-phonon scattering through vertex correction.
\citet{ferrante2018} calculated the self-energy of the zone-center optical phonons of graphene based on both intra-band and inter-band electronic transitions, accounting for the finite electron linewidths of both linear bands in the case of the latter; see Eq.~(2) therein, which is reasonable and yet intuitive. Moreover, Eq.~(2) of \citet{ferrante2018} employs double integration over the energies of the two electronic states, the integration over $z$ and $z'$ in Eq.~(2) of \citet{ferrante2018}, involved in the phonon-induced scattering.

This paper presents the theory for non-adiabatic phonon self-energy arising from coupling to electrons with finite linewidths, accounting for both intra-band and inter-band transitions and the linewidths of both bands in the latter, based on the electron Green's functions. The final form of our theory [Eqs.~\eqref{eq:Pi_omega3} and~\eqref{eq:Kramers_Kronig}] does not have double energy integration and is suitable for a straightforward implementation in software packages on electron-phonon interactions.

The remainder of the paper is organized as follows. In Sec.~\ref{sec:methods}, we present the theory and method for first-principles calculations of non-adiabatic phonon self-energy, and in Sec.~\ref{sec:results} we apply the method to graphene at low temperatures. We conclude in Sec.~\ref{sec:conclusion}. The derivations of some of the key equations in Sec.~\ref{sec:methods} are presented in Sec.~\ref{sec:appendix}.

\section{Method}
\label{sec:methods}
Using the spectral representation of the Matsubara dressed electron Green functions~\cite{MahanBook}, we can write the imaginary part of the non-adiabatic phonon self-energy as
\begin{widetext}
\begin{eqnarray}
\label{eq:Pi_spectral}
&&{\rm Im\,} \Pi^{\rm NA}_{\qnu}(i\omega_m) = {\rm Im\,}\, \frac{1}{\beta}\,\sum_{mn}\,\int \frac{d{\bf k}}{\Omega_{\rm BZ}} 
|g_{mn\nu}({\bf k},{\bf q})|^2\,
\sum_{i p_n} G_{\nk}(ip_n)\, G_{\mkq}(ip_n+i\omega_m)\nnnl
&=&\int d\varepsilon\,\delta_{\nk} (\varepsilon)\, \int d\varepsilon'\,\delta_{\mkq} (\varepsilon')\,
{\rm Im\,}\, \frac{1}{\beta}\,\sum_{mn}\,\int \frac{d{\bf k}}{\Omega_{\rm BZ}} 
|g_{mn\nu}({\bf k},{\bf q})|^2\,
\sum_{i p_n}\,
\frac{1}{ip_n - \varepsilon+\mu}\,\frac{1}{ip_n+i\omega_m - \varepsilon'+\mu}\nnnl
&=&\int d\varepsilon\,\delta_{\nk} (\varepsilon)\, \int d\varepsilon'\,\delta_{\mkq} (\varepsilon')\,
{\rm Im\,}\, \frac{1}{\beta}\,\sum_{mn}\,\int \frac{d{\bf k}}{\Omega_{\rm BZ}} 
|g_{mn\nu}({\bf k},{\bf q})|^2\,
\frac{1}{\varepsilon-\varepsilon'+i\omega_m}\,
\sum_{i p_n}\,\left(
\frac{1}{ip_n - \varepsilon+\mu}-\frac{1}{ip_n+i\omega_m - \varepsilon'+\mu}\right)\nnnl
&=&\int d\varepsilon\,\delta_{\nk} (\varepsilon)\, \int d\varepsilon'\,\delta_{\mkq} (\varepsilon')\,
{\rm Im\,}\, \sum_{mn}\,\int \frac{d{\bf k}}{\Omega_{\rm BZ}} 
|g_{mn\nu}({\bf k},{\bf q})|^2\,
\frac{f(\varepsilon)-f(\varepsilon')}{\varepsilon-\varepsilon'+i\omega_m}\,,\nnnl
\end{eqnarray}
\end{widetext}
where $\beta=1/k_{\rm B}T$ and $\omega_m=2\pi m/\beta$ and $p_n=\pi (2n+1)/\beta$ for integers $m$ and $n$ are the bosonic and fermionic Matsubara frequencies, respectively.
In the last equality, the series expansion of the Fermi-Dirac distribution~\cite{MahanBook}
\begin{equation}
\label{eq:f_Matsubara}
f(z)=\frac{1}{2} - \frac{1}{\beta} \sum_{n=-\infty}^{\infty} \frac{1}{z - \mu - ip_n}\,.
\end{equation}
was used.
Also,
\begin{equation}
\label{eq:delta_mk}
\delta_{\nk} (\varepsilon) = \frac{1}{\pi} \frac{\Sigma{''}_{\nk}}{(\varepsilon-\varepsilon_{\nk})^2+\Sigma{''}_{\nk}^2}
\end{equation}
is the electron spectral function, where
\begin{eqnarray}
\label{eq:ImSigma}
\Sigma{''}_{\nk}&\equiv&-{\rm Im\,}\Sigma_{\nk}(\varepsilon_{\nk})>0
\end{eqnarray}
is the magnitude of the imaginary part of the retarded electron self-energy.
We note that
\begin{equation}
\label{eq:delta_limit}
\lim_{\Sigma{''}_{\nk}\to0^+}\delta_{\nk} (\varepsilon) = \delta(\varepsilon-\varepsilon_{\nk})\,.
\end{equation}
{ Non-collinear spinor wavefunctions were assumed. In case the system is a collinear magnet, one can denote the composite band index $m$ or $n$ in Eq.~\eqref{eq:Pi_spectral} by the (up or down) spin index and the band index at a given spin.}
For non-magnetic, non-spinor calculations, we can multiply Eq.~\eqref{eq:Pi_spectral} by 2 and confine the band summation to bands with one kind of spin. In other cases, this simplification cannot be used.

To obtain the retarded phonon self-energy, we replace $i\omega_m$ in Eq.~\eqref{eq:Pi_spectral} with $\omega+i0^+$.
{ Using $\frac{1}{x+i0^+}=P\left(\frac{1}{x}\right)-i\pi\delta(x)$, we obtain}
\begin{widetext}
\begin{eqnarray}
\label{eq:Pi_omega}
&&{\rm Im\,} \Pi^{\rm NA}_{\qnu}(\omega+i0^+)%\nnnl
%&=&
=-\pi\,\sum_{mn}\,\int \frac{d{\bf k}}{\Omega_{\rm BZ}} 
|g_{mn\nu}({\bf k},{\bf q})|^2\,
\int d\varepsilon\,\delta_{\nk} (\varepsilon)\, \delta_{\mkq} (\varepsilon+\omega)\,
\left[f(\varepsilon)-f(\varepsilon+\omega)\right]\nnnl
&=&-\frac{1}{\pi}\,\sum_{mn}\,\int \frac{d{\bf k}}{\Omega_{\rm BZ}} 
|g_{mn\nu}({\bf k},{\bf q})|^2\,
\int d\varepsilon\,\frac{\Sigma{''}_{\nk}}{(\varepsilon-\varepsilon_{\nk})^2+\Sigma{''}_{\nk}^2}\,
\frac{\Sigma{''}_{\mkq}}{(\varepsilon+\omega-\varepsilon_{\mkq})^2+\Sigma{''}_{\mkq}^2}
\left[f(\varepsilon)-f(\varepsilon+\omega)\right]\,.\nnnl
\end{eqnarray}
\end{widetext}

Equation~\eqref{eq:Pi_omega} can be intuitively interpreted as the usual linewidth formula [Eq.~\eqref{eq:Pi_omega_non_adiabatic_Allen}, but replacing $\omega_\qnu$ with $\omega$, or, equivalently, Eq.~\eqref{eq:Pi_omega_limit3}] broadened by the electron linewidths.
{ We note that the imaginary part of Eq.~(2) of \citet{ferrante2018}, after the integration over $z$ is performed, is equivalent to Eq.~\eqref{eq:Pi_omega}; \citet{ferrante2018} arrived at their Eq.~(2) from physical intuition, rather than from a rigorous derivation. Our work proves that this result is rigorously correct using the Green's function method.}
In the limit the electron linewidths vanish, Eq.~\eqref{eq:Pi_omega} reduces to the usual result for the case with zero electron linewidths [Eq.~\eqref{eq:Pi_omega_limit3}] as we can easily see with the help of Eq.~\eqref{eq:delta_limit}.

We will now convert Eq.~\eqref{eq:Pi_omega} to a form that does not have energy integration. Following Ref.~\citealp{2022Pickem_PRB_Digamma} and the supplementary information of Ref.~\citealp{2024LihmSCEL}, we will use the digamma function,
\begin{equation}
\label{eq:digamma}
\psi(z)=-\gamma + \sum_{n=1}^\infty \left( \frac{1}{n}-\frac{1}{n+z-1} \right)\,,
\end{equation}
where $\gamma$ is the Euler-Mascheroni constant.

After a lengthy but straightforward derivation (see Sec.~\ref{sec:main_derivation}), we obtain
\begin{widetext}
\begin{eqnarray}
\label{eq:Pi_omega3}
&&{\rm Im\,} \Pi^{\rm NA}_{\qnu}(\omega+i0^+)=-\frac{1}{2\pi}\,\sum_{mn}\,\int \frac{d{\bf k}}{\Omega_{\rm BZ}} 
|g_{mn\nu}({\bf k},{\bf q})|^2\,\times\nnnl
&&{\rm Re\,}\left\{ \frac{-\psi\left[\tfrac{1}{2} - \tfrac{\beta}{2\pi i}\,(\varepsilon_{\nk}-\mu-i\Sigma{''}_{\nk})\right] + \psi\left[\tfrac{1}{2} - \tfrac{\beta}{2\pi i}\,(-\omega+\varepsilon_{\mkq}-\mu-i\Sigma{''}_{\mkq})\right]}{\varepsilon_\nk + \omega - \varepsilon_\mkq-i(\Sigma{''}_\nk-\Sigma{''}_\mkq)} \right.
\nnnl
&&+\frac{\psi\left[\tfrac{1}{2} - \tfrac{\beta}{2\pi i}\,(\varepsilon_{\nk}-\mu-i\Sigma{''}_{\nk})\right]  - \psi\left[\tfrac{1}{2} + \tfrac{\beta}{2\pi i}\,(-\omega+\varepsilon_{\mkq}-\mu+i\Sigma{''}_{\mkq})\right]}{\varepsilon_\nk + \omega - \varepsilon_\mkq-i(\Sigma{''}_\nk+\Sigma{''}_\mkq)}
\nnnl
&&+\frac{\psi\left[\tfrac{1}{2} - \tfrac{\beta}{2\pi i}\,(\varepsilon_{\nk}+\omega-\mu-i\Sigma{''}_{\nk})\right]-\psi\left[\tfrac{1}{2} - \tfrac{\beta}{2\pi i}\,(\varepsilon_{\mkq}-\mu-i\Sigma{''}_{\mkq})\right] }{\varepsilon_\nk + \omega - \varepsilon_\mkq-i(\Sigma{''}_\nk-\Sigma{''}_\mkq)}
\nnnl
&&+\left.\frac{- \psi\left[\tfrac{1}{2} + \tfrac{\beta}{2\pi i}\,(\varepsilon_{\nk}+\omega-\mu+i\Sigma{''}_{\nk})\right]+\psi\left[\tfrac{1}{2} - \tfrac{\beta}{2\pi i}\,(\varepsilon_{\mkq}-\mu-i\Sigma{''}_{\mkq})\right] }{\varepsilon_\nk + \omega - \varepsilon_\mkq+i(\Sigma{''}_\nk+\Sigma{''}_\mkq)}\right\}\,.
\nnnl
\end{eqnarray}
\end{widetext}
Equation~\eqref{eq:Pi_omega3} is the main result of this paper and is suitable for first-principles calculations.
We note that in practice the digamma function is not evaluated according to its definition in Eq.~\eqref{eq:digamma} in numerical computations; special functions libraries allow the efficient evaluation of this function.
{The digamma function is often evaluated with a mixture of series expansion, asymptotic approximations, and reflection formulas depending on the argument. See, for example, \citet{1976Bernardo}.}

The real part of the non-adiabatic phonon self-energy [see Eq.~(145) of Ref.~\citealp{2017Giustino_RMP_ElPh}] can be obtained from the Kramers-Kronig relation:
\begin{eqnarray}
\label{eq:Kramers_Kronig}
&&{\rm Re\,} \Pi^{\rm NA}_{\qnu}(\omega+i0^+)=\frac{1}{\pi}\,\int d\omega'\,\frac{{\rm Im\,} \Pi^{\rm NA}_{\qnu}(\omega'+i0^+)}{\omega'-\omega}\nnnl
&&-\sum_{mn}\,\int \frac{d{\bf k}}{\Omega_{\rm BZ}} 
|g_{mn\nu}({\bf k},{\bf q})|^2\,\frac{f(\varepsilon_\nk)-f(\varepsilon_\mkq)}{\varepsilon_\nk-\varepsilon_\mkq}\,.\nnnl
\end{eqnarray}

Regarding Eq.~\eqref{eq:Pi_omega3}, we first note that if $\varepsilon_\nk+\omega=\varepsilon_\mkq$ and $\Sigma{''}_\nk=\Sigma{''}_\mkq$, both the denominator and the numerator of the first and third fractions vanish. However, these fractions are well-defined if the denominator is not precisely zero. In practice, we add a tiny positive number to the largest among $\Sigma{''}_\nk$ and $\Sigma{''}_\mkq$ to avoid numerical problems. Secondly, Eq.~\eqref{eq:Pi_omega3} naturally allows the intra-band contribution, i.\,e.\,, the case with $m=n$, which was forbidden for zone-center optical phonons if zero electron linewidths were used.

We now discuss the limiting cases of Eq.~\eqref{eq:Pi_omega3}. See Sec.~\ref{sec:limiting_cases} for the derivation.
%First, the $\Sigma{''}_\nk\to0^+$ limit of Eq.~\eqref{eq:Pi_omega3} is
The limit $\Sigma{''}_\nk\to0^+$ and the limit $\Sigma{''}_\mkq\to0^+$ of Eq.~\eqref{eq:Pi_omega3} are, respectively,
\begin{widetext}
\begin{subequations}
\begin{eqnarray}
\label{eq:Pi_omega_limit1_short}
&&{\rm Im\,} \Pi^{\rm NA}_{\qnu}(\omega+i0^+)=-\pi\,\sum_{mn}\,\int \frac{d{\bf k}}{\Omega_{\rm BZ}} 
|g_{mn\nu}({\bf k},{\bf q})|^2\,\left[f(\varepsilon_\nk)-f(\varepsilon_\nk+\omega)\right]\,
\delta_{\mkq} (\varepsilon_\nk+\omega)\,.
\end{eqnarray}
and
\begin{eqnarray}
\label{eq:Pi_omega_limit2_short}
{\rm Im\,} \Pi^{\rm NA}_{\qnu}(\omega+i0^+)=-\pi\,\sum_{mn}\,\int \frac{d{\bf k}}{\Omega_{\rm BZ}} 
|g_{mn\nu}({\bf k},{\bf q})|^2\,\left[f(-\omega+\varepsilon_\mkq)-f(\varepsilon_\mkq)\right]\,
\delta_{\nk} (-\omega+\varepsilon_\mkq)\,.
\end{eqnarray}
\end{subequations}
\end{widetext}

Finally, the $\Sigma{''}_\mkq\to0^+$ limit of Eq.~\eqref{eq:Pi_omega_limit1_short}
and the $\Sigma{''}_\nk\to0^+$ limit of Eq.~\eqref{eq:Pi_omega_limit2_short} both are
\begin{widetext}
\begin{equation}
\label{eq:Pi_omega_limit3}
{\rm Im\,} \Pi^{\rm NA}_{\qnu}(\omega+i0^+)=-\pi\,\sum_{mn}\,\int \frac{d{\bf k}}{\Omega_{\rm BZ}} 
|g_{mn\nu}({\bf k},{\bf q})|^2\,\left[f(\varepsilon_\nk)-f(\varepsilon_\mkq)\right]\,
\delta (\varepsilon_\nk+\omega-\varepsilon_\mkq)\,\,,
\end{equation}
\end{widetext}
which is the correct result for infinitesimal electron linewidths~\cite{2017Giustino_RMP_ElPh} and is equivalent to Eq.~\eqref{eq:Pi_omega_non_adiabatic_Allen} for $\omega=\omega_\qnu$.
The factor 2 difference between Eq.~\eqref{eq:Pi_omega_limit3} and Eqs.~(145) and~(146) of Ref.~\citealp{2017Giustino_RMP_ElPh} is that the latter considered non-magnetic, non-spinor cases and confined the band summation to only over the bands of one kind of spin.

\section{Linewidth of the $E_{\rm 2g}$ optical phonon at $\Gamma$ of graphene}
\label{sec:results}

As a proof of principle, we will apply Eq.~\eqref{eq:Pi_omega3} to graphene and calculate the double-degenerate zone-center $E_{\rm 2g}$ optical phonon linewidth arising from electron-phonon interactions:
\begin{equation}
\label{eq:Gamma}
\Gamma_\qnu = -2\,{\rm Im\,} \Pi^{\rm NA}_{\qnu}(\omega_\qnu+i0^+)\,,
\end{equation}
where ${\bf q}=\Gamma$ and $\nu$ is the longitudinal optical or transverse optical branch.

The electronic band velocity, the energy of the zone-center $E_{\rm 2g}$ optical phonon, and electron-phonon coupling constants were taken from the model in Tab.~I of Ref.~\citealp{2014Park_NL_graphene_resistivity}; this model was made from first-principles calculations and reproduced the experimentally measured intrinsic electrical resistivity of graphene~\cite{2014Park_NL_graphene_resistivity} and was also applied to the calculation of the electronic thermal conductivity of graphene~\cite{2016Kim_NL_graphene_thermal_conductivity}. Following this model, we included two linear energy bands with band indices $+1$ and $-1$ (for both $m$ and $n$), whose energy is above and below the Dirac point energy, respectively.
For the numerical calculation of wavevector integration, we included states with energies between $-4$~eV and $4$~eV, where the Dirac point energy is set to zero. The momentum integration in Eq.~\eqref{eq:Pi_omega3} along the radial direction from any one of the two Dirac points was performed with the energy range divided into 20,000 uniformly sampled points, and that along the circumferential direction was performed analytically.

If we account for the spin and valley degrees of freedom, the phonon linewidth [Eq.~\eqref{eq:Gamma}] of the degenerate zone-center $E_{\rm 2g}$ optical phonons of graphene reduces to
\begin{widetext}
\begin{eqnarray}
\label{eq:Gamma_graphene}
&&\Gamma_{E_{\rm 2g}} = \frac{3\sqrt{3}}{\pi^2}\left(\frac{b\,\bar{g}}{\hbar\,v_{\rm F}}\right)^2 \sum_{m,\,n=\,\{-1,+1\}}\,\int_0^{4~{\rm eV}} d\varepsilon\,\varepsilon\,{\rm Re\,}\left\{...\right\}\,,
\end{eqnarray}
\end{widetext}
where $b$ is the carbon-carbon bond length in graphene, $v_{\rm F}$ is the band velocity around the Dirac point, and $\bar{g}$ is the angle-averaged electron-phonon coupling constant and is given by $\bar{g}=\frac{3}{2}\,\eta\,\sqrt{\frac{\hbar}{2M_{\rm C}\omega_\Gamma^{{\rm E}_{2g}}}}$ using the notation in Tab.~I of Ref.~\citealp{2014Park_NL_graphene_resistivity}.
{Here, $M_\textrm{C}$ is the mass of a carbon atom, $\hbar\,\omega_\Gamma^{\textrm{E}_{2g}}=0.2$~eV is the energy of the zone-center $E_\textrm{2g}$ optical phonon (note the difference of factor $\hbar$ in the definition of $\omega_\qnu$ between the current paper and Ref.~21), and $\eta=\frac{2\hbar v_\textrm{F}}{3b^2}\left(1-\frac{b}{v_\textrm{F}}\frac{dv_\textrm{F}}{db}\right)=4.75$~eV$\cdot$\AA$^{-1}$ is the electron-phonon coupling strength, where $b=1.405$~\AA\ is the carbon-carbon bond length and $v_\textrm{F}=0.866\times10^6$~m/s is the electron band velocity of graphene at the Dirac point.
See Sec.~\ref{sec:graphene_derivation} for the derivation of Eq.~\eqref{eq:Gamma_graphene}.}
Also, ${\rm Re\,}\left\{...\right\}$ in Eq.~\eqref{eq:Gamma_graphene} is obtained from ${\rm Re\,}\left\{...\right\}$ in Eq.~\eqref{eq:Pi_omega3} by replacing $\omega$ with $\omega_\qnu=0.2$~eV, $\varepsilon_\nk$ and $\varepsilon_\mkq$ with $m\,\varepsilon$ and $n\,\varepsilon$, respectively, and $\Sigma{''}_\mkq$ and $\Sigma{''}_\nk$ with $\Sigma{''}(m\,\varepsilon)$ and $\Sigma{''}(n\,\varepsilon)$, respectively. Here, the function $\Sigma{''}(\varepsilon)$ is set to a good approximation to give the magnitude of the imaginary part of the experimentally measured electron self-energy~\cite{2007Boswick_NatPhys_graphene_ARPES}, which is in good agreement with the first-principles calculations~\cite{2007Park_PRL_graphene_eph, 2009Park_PRL_graphene_linewidth, 2009Park_NL_graphene_ARPES}:
\begin{eqnarray}
\label{eq:ImSigma_graphene}
&&\Sigma{''}(\varepsilon) =\nnnl
&&
\begin{cases}
0.054\,|\varepsilon-0.2~\text{eV}|+0.01~\text{eV}&\text{if}\,\varepsilon >\mu+0.2~\text{eV}
\\
0.054\,|\varepsilon+0.2~\text{eV}|+0.01~\text{eV} & \text{if}\,\varepsilon<\mu-0.2~\text{eV}
\\
0.01~\text{eV} &\text{otherwise}\,.
\end{cases}
\end{eqnarray}
Note that the calculated phonon linewidths are not very sensitive to $\Sigma{''}(\varepsilon)$ as long as it approximates the imaginary part of the measured electron self-energy. Note also that the electron linewidths depend appreciably on the quality of the sample~\cite{2007Yan_PRL_graphene_phonon_linewidth,PhysRevB.91.205413,PhysRevB.107.075420}. Therefore, the phonon linewidths may also depend on the sample quality.

\begin{figure*}[tbp]
\centering
\includegraphics[width=1.0\textwidth]{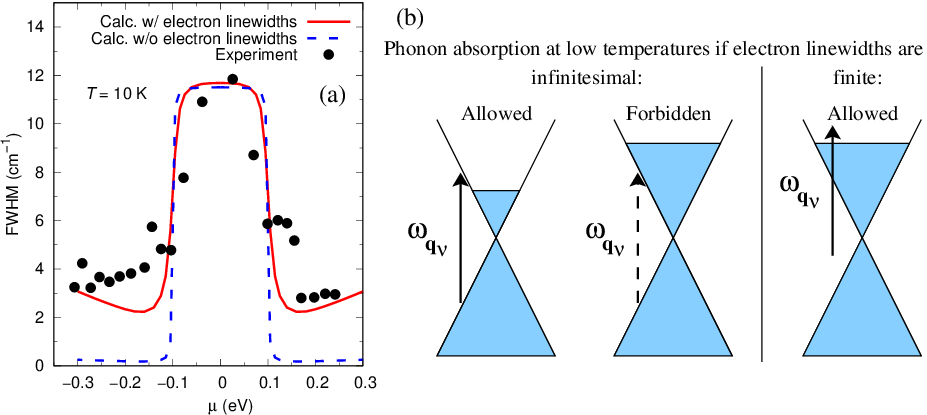}
\caption{
(a) Linewidth [full width at half maximum (FWHM)] of the zone-center $E_{\rm 2g}$ optical phonons of graphene vs. chemical potential at $T=10$~K. Theoretical results were obtained according to Eq.~\eqref{eq:Gamma_graphene} with finite electron linewidths [Eq.~\eqref{eq:ImSigma_graphene}] (solid or red curve) or with infinitesimal electron linewidths - we assumed 0.001~eV linewidths - (dashed or blue curve) from Eq.~\eqref{eq:Pi_omega3}. We took the experimental data from Ref.~\citealp{2007Yan_PRL_graphene_phonon_linewidth} and shifted them by $(-3)$~cm$^{-1}$ to account for other phonon scattering mechanisms than electron-phonon interactions (see text).
(b) Schematic showing that a zone-center $E_{\rm 2g}$ optical phonon in graphene can decay through electronic transitions at low temperatures only if $|\mu|<\omega_\qnu/2$ if the electron linewidths are infinitesimal. { However, if finite electron linewidths are considered, the optical phonon can decay even when $|\mu|>\omega_\qnu/2$.}
}
\label{fig:graphene_experiment}
\end{figure*}

Figure~\ref{fig:graphene_experiment}(a) shows that, at low temperatures, the linewidth of the degenerate zone-center $E_{\rm 2g}$ optical phonons of graphene calculated with infinitesimal electron linewidths is constant if the chemical potential is between $-0.1$~eV and $+0.1$~eV from the Dirac point energy and is zero outside this window. The reason for this behavior is illustrated in Fig.~\ref{fig:graphene_experiment}(b). At low temperatures, if $|\mu|<\omega_\qnu/2=0.1$~eV, a phonon with energy $\omega_\qnu=0.2$~eV can decay by an electronic transition; however, if $|\mu|>\omega_\qnu/2$, this transition is Pauli-blocked, and the phonon is not absorbed by vertical electronic transitions. Our calculated results for zero electron linewidths are in good agreement with previous first-principles calculations~\cite{2006Lazzeri_PRL_graphene_Kohn_anomaly,2008Park_NL_graphene_eph}. However, the results calculated with finite electron linewidths [Eqs.~\eqref{eq:Gamma_graphene} and~\eqref{eq:ImSigma_graphene}] are different from these results: Even in graphene with high charge density, the scattering rate is finite, because energy conservation [Eq.~\eqref{eq:energy_conservation}] with sharply defined electronic energies is no longer required; hence, all electronic states contribute to the linewidth of the $E_{\rm 2g}$ optical phonons.
{ This mechanism is also explained in Fig.~\ref{fig:graphene_experiment}(b)}.

\begin{figure}[tbp]
\centering
\includegraphics[width=0.98\columnwidth]{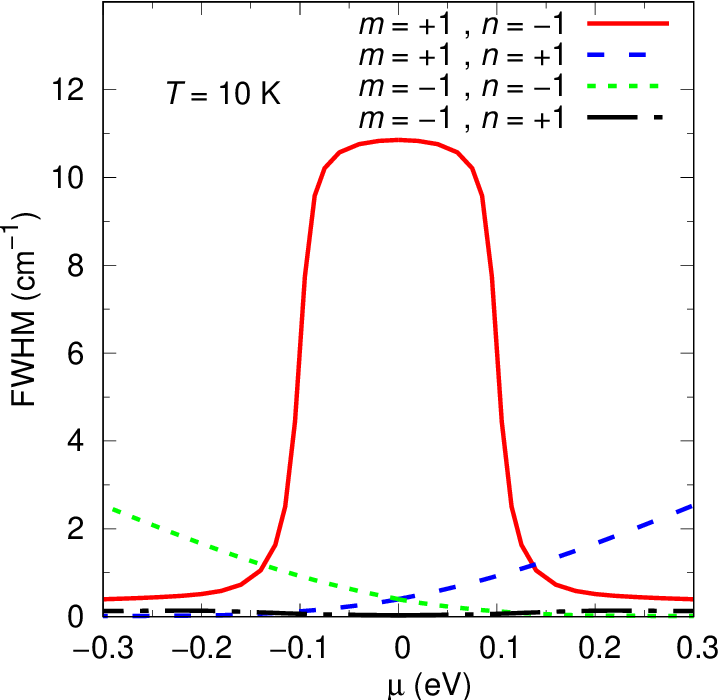}
\caption{
Band-resolved phonon linewidth [each summand of Eq.~\eqref{eq:Gamma_graphene}] calculated with finite electron linewidths. $m$ and $n$ indicate, respectively, the electronic bands {\it after} and {\it before} absorbing a zone-center $E_{\rm 2g}$ optical phonon of graphene. All four contributions amount to the sold or red curve in Fig.~\ref{fig:graphene_experiment}.
}
\label{fig:graphene_band_resolution}
\end{figure}

To compare the calculated phonon linewidths with the experimental data in~\citet{2007Yan_PRL_graphene_phonon_linewidth}, we shifted the experimental data by $-3$~cm$^{-1}$ to match the phonon linewidth obtained from computations at zero charge density because there are other scattering mechanisms, which are not very sensitive to the charge density such as phonon-phonon scattering, for phonon linewidths measured from experiments. We note that~\citet{2007Bonini_PRL_graphene_phonon} have shown that three-phonon scattering processes at low temperatures add 1.65~cm$^{-1}$ to the total phonon linewidth. Therefore, other scattering mechanisms, such as phonon-impurity scattering and higher-order scattering processes, are responsible for the remaining shift in the experimental data, $3-1.65=1.35$~cm$^{-1}$.
Notably, the experimental data on the linewidths vs. chemical potential are much more in agreement with the calculations obtained with finite electron linewidths. Our calculations with finite electronic linewidths even reproduce the {\it increase} in the measured linewidth with chemical potential at higher charge density, which is due to intra-band electronic transitions as we explain below and show in Fig.~\ref{fig:graphene_band_resolution}.

To better understand the computational results, we divide the total phonon linewidth into contributions from different electronic transitions, i.\,e.\,, with different $m$ and $n$ pairs. The results are shown in Fig.~\ref{fig:graphene_band_resolution}. The $n=-1$ to $m=+1$ transition, i.\,e.\,, the process in which a negative energy electronic state absorbs a phonon and makes a transition to a positive energy electronic state, is by far dominant in the phonon linewidths. Also, this contribution does not disappear even if $|\mu|>\omega_\qnu/2$. However, the most interesting result is that the intra-band contributions, that is, processes involving the electronic transition in the same band ($m=n$), increase with $|\mu|$ and are dominant at high charge densities. This behavior is also clearly observed in the Raman experiment by \citet{PhysRevB.91.205413} [see Fig.~9(a) thereof].
{ As long as the intra-band electron-phonon coupling matrix elements for the zone-center $E_\textrm{2g}$ phonon are non-zero, and the energy conservation condition can be relaxed, the zone-center $E_\textrm{2g}$ phonon can decay through intra-band electron scattering processes. Consider the electron-phonon interaction Hamiltonian within the second quantization formalism: Destruction of an electronic state and its subsequent creation is an allowed operation. Relaxation of the energy conservation can be intuitively understood as follows: Since there are uncertainties in the energies of the initial and final electronic states, we can find pairs of electronic states whose energy difference matches the energy of the $E_\textrm{2g}$ optical phonon.}

\begin{figure*}[tbp]
\centering
\includegraphics[width=1.0\textwidth]{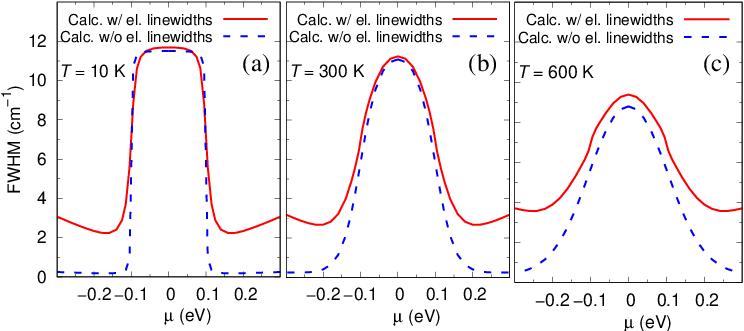}
\caption{
Linewidth of the zone-center $E_{\rm 2g}$ optical phonons of graphene at different temperatures calculated according to Eq.~\eqref{eq:Gamma_graphene} with finite electron linewidths [Eqs.~\eqref{eq:ImSigma_graphene}] and without finite electron linewidths - we assumed 0.001~eV linewidths. Here, we did not consider the increase in the electron linewidth with temperature { because that is not the subject of the paper. If that effect is considered, phonon linewidths may further increase at higher temperatures.}
}
\label{fig:temp_variations}
\end{figure*}

\begin{figure*}[tbp]
\centering
\includegraphics[width=1.0\textwidth]{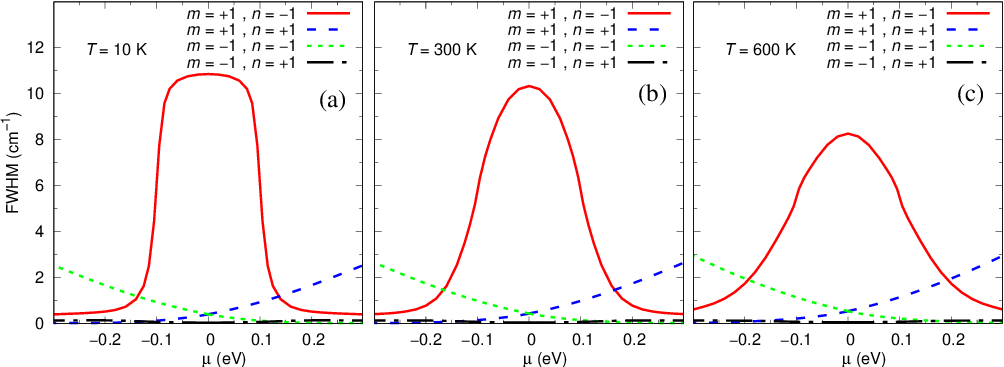}
\caption{
{
The same quantities as in Fig.~\ref{fig:graphene_band_resolution} at different temperatures.}
}
\label{fig:temp_band_resolution}
\end{figure*}

Figure~\ref{fig:temp_variations} shows the calculated phonon linewidths at different temperatures. The linewidths vs. chemical potential curves obtained with infinitesimal electron linewidths become smoother as the temperature increases. The results obtained with finite electron linewidths show additional features and, in general, vary more smoothly with $|\mu|$, and the phonon linewidth increases with $|\mu|$ in the high-charge-density regime.
{ Interestingly, even if we assume zero electron linewidths, the zone-center $E_\textrm{2g}$ phonon linewidths of low-doped graphene arising from electron-phonon interactions \textit{decrease} with temperature (see the dashed blue curves in Fig.~\ref{fig:temp_variations} and also the upper panel of Fig.~3 of \citet{2006Lazzeri_PRL_graphene_Kohn_anomaly}); at higher temperatures, the low-energy (initial) electronic state ($\varepsilon=-\omega_\qnu/2$) is less occupied, and the high-energy electronic state ($\varepsilon=+\omega_\qnu/2$) is less empty, which results in blocking the available phase space for the phonon decay.
In passing, we note that the electron linewidths will increase with temperature, which is not considered in the calculation presented in this work because it is not the subject of this study. If this effect is considered, the phonon linewidths will increase further with temperature.

Figure~\ref{fig:temp_band_resolution} shows band-resolved contributions to phonon linewidths at different temperatures.

Finally, one can think of the self-consistent determination of both electron and phonon linewidths. In fact, the theory of computing electron linewidths self-consistently assuming zero phonon linewidths was developed by \citet{2024LihmSCEL}, and the theory of computing electron linewidths, assuming finite phonon linewidths but zero electron linewidths was developed by \citet{Lihm2024PlPh} [see Eq. (S71) therein]. However, the theory of computing electron linewidths assuming both finite electron linewidths and finite phonon linewidths has yet to be developed. In addition to this problem, there are other scattering mechanisms for electrons and those for phonons beyond the electron-phonon interactions that need to be taken into account, which, in principle, can be done, for the self-consistent computation of electron and phonon linewidths.
}

\section{Conclusion}
\label{sec:conclusion}

In this paper, we presented the theory for calculating non-adiabatic phonon self-energy arising from the coupling to electrons with finite linewidths.
The theory naturally introduces the contribution to phonon linewidth from intra-band electronic transitions for all electronic states, not just the ones satisfying the energy conservation condition upon phonon absorption.

As a proof of principle, we demonstrated that the measured zone-center $E_{\rm 2g}$ optical phonon linewidth vs. chemical potential behavior of graphene at low temperatures can be quantitatively explained only if the finite electron linewidths are considered.
Remarkably, we find that the intra-band electronic transitions, which are forbidden in this example if infinitesimal electronic linewidths are used, contribute significantly to the 
self-energy of the zone-center $E_{\rm 2g}$ optical phonons of graphene.

Our formalism can be easily implemented in existing software packages for computing electron-phonon interactions from first principles, such as \textsc{EPW}~\cite{2016PonceEPW}, \textsc{Abinit}~\cite{Gonze2020}, \textsc{Perturbo}~\cite{2021ZhouPerturbo}, \textsc{Phoebe}~\cite{2022CepellottiPhoebe}, \textsc{elphbolt}~\cite{2022Nakib_elphbolt}, and \textsc{EPIq}~\cite{2024MariniEPIq}, and \textsc{ElectronPhonon.jl}~\cite{Lihm2024NLHE,EPjl}.
We expect that the theory will be necessary to interpret experimental data on materials whose phonon energy scale is non-negligible compared to the energy scale of low-energy electronic transitions, which are materials of contemporary interest.

\begin{acknowledgments}
The author thanks Fabio Caruso for drawing the author's attention to this problem and for helpful discussions, Francesco Mauri and Dino Novko for helpful discussions and comments on the preprint of this manuscript, Jae-Mo Lihm and Samuel Ponc\'e for helpful discussions through collaborations, and Samuel Ponc\'e, Xavier Gonze, Andrea Marini, Fabio Caruso, and Matthieu Vestraete, the organizers of the European Theoretical Spectroscopy Facility (ETSF) Electron-phonon collaboration team workshop held in Universit\'e catholique de Louvain, Louvain-la-Neuve, Belgium, on September 23--25, 2024, during which this study was initiated. This work was supported by the Korean NRF No-2023R1A2C1007297. Computational resources were provided by the KISTI Supercomputing Center (Grant No. KSC-2023-CRE-0533).
\end{acknowledgments}

\section{Appendix}
\label{sec:appendix}

\subsection{Derivation of Eq.~\eqref{eq:Pi_omega3}}
\label{sec:main_derivation}
We can rewrite Eq.~\eqref{eq:Pi_omega} as
\begin{widetext}
\begin{eqnarray}
\label{eq:Pi_omega2}
&&{\rm Im\,} \Pi^{\rm NA}_{\qnu}(\omega+i0^+)=\frac{1}{\pi\,\beta}\,\sum_{mn}\,\int \frac{d{\bf k}}{\Omega_{\rm BZ}} 
|g_{mn\nu}({\bf k},{\bf q})|^2\times\nnnl
&&\left[\sum_{l=-\infty}^{\infty}\int d\varepsilon\,\frac{1}{\varepsilon-\mu-ip_l}\,\frac{\Sigma{''}_{\nk}}{(\varepsilon- \varepsilon_{\nk})^2+\Sigma{''}_{\nk}^2}\,
\frac{\Sigma{''}_{\mkq}}{(\varepsilon+\omega-\varepsilon_{\mkq})^2+\Sigma{''}_{\mkq}^2}
-\left( \nk \xleftrightarrow{} \mkq,\, \omega \xleftrightarrow{} -\omega \right)\right]\nnnl
\end{eqnarray}
\end{widetext}
Then, we can do the energy integration as follows:
\begin{widetext}
\begin{eqnarray}
\label{eq:energy_integral}
&&\sum_{l=-\infty}^{\infty}\int d\varepsilon\,\frac{1}{\varepsilon-\mu-ip_l}\,\frac{\Sigma{''}_{\nk}}{(\varepsilon- \varepsilon_{\nk})^2+\Sigma{''}_{\nk}^2}\,
\frac{\Sigma{''}_{\mkq}}{(\varepsilon+\omega-\varepsilon_{\mkq})^2+\Sigma{''}_{\mkq}^2}
=-\frac{1}{4}\times
\nnnl
&&\sum_{l=-\infty}^{\infty}\int d\varepsilon\,\frac{1}{\varepsilon-\mu-ip_l}\,
\left(
\frac{1}{\varepsilon- \varepsilon_{\nk}+i\Sigma{''}_{\nk}}
-\frac{1}{\varepsilon- \varepsilon_{\nk}-i\Sigma{''}_{\nk}}
\right)\,
\left(\frac{1}{\varepsilon+\omega-\varepsilon_{\mkq}+i\Sigma{''}_{\mkq}}-\frac{1}{\varepsilon+\omega-\varepsilon_{\mkq}-i\Sigma{''}_{\mkq}}\right).
\nnnl
\end{eqnarray}
\end{widetext}
Now, the right-hand side of Eq.~\eqref{eq:energy_integral} can be split into four terms. Note that $\Sigma{''}_\nk$ and $\Sigma{''}_\mkq$ are positive [Eq.~\eqref{eq:ImSigma}].

We will make use of the following identity frequently:
\begin{eqnarray}
\label{eq:Matsubara_sum_digamma}
    &&\sum_{l=0}^{\infty} \frac{1}{(ip_l + a)(ip_l + b)}\nnnl
    &=& \frac{\beta^2}{(2\pi i)^2} \sum_{l=0}^{\infty} \frac{1}{(l + \tfrac{1}{2} + \tfrac{\beta}{2\pi i}\, a)(l + \tfrac{1}{2} + \tfrac{\beta}{2\pi i}\,b )}\nnnl
    &=& \frac{\beta}{2\pi i} \frac{\psi(\tfrac{1}{2} + \tfrac{\beta}{2\pi i}\,a) - \psi(\tfrac{1}{2} + \tfrac{\beta}{2\pi i}\,b)}{a-b}\,,
\end{eqnarray}
which follows from
\begin{equation}
\label{eq:digamma_equality}
    \sum_{l=0}^{\infty} \frac{1}{(l+a)(l+b)} = \frac{\psi(a) - \psi(b)}{a-b}\,,
\end{equation}
which again follows from the definition of the digamma function in Eq.~\eqref{eq:digamma}.
Equations~\eqref{eq:Matsubara_sum_digamma} and~\eqref{eq:digamma_equality} correspond to Eqs.~(S44) and~(S45) of Ref.~\citealp{2024LihmSCEL}.

We evaluate the first part in the second line of Eq.~\eqref{eq:energy_integral} by choosing an infinitely large upper semicircle as its contour:
\begin{widetext}
\begin{eqnarray}
\label{eq:energy_integral1}
&&\sum_{l=-\infty}^{\infty}\int d\varepsilon\,\frac{1}{\varepsilon-\mu-ip_l}\,
\frac{1}{\varepsilon- \varepsilon_{\nk}+i\Sigma{''}_{\nk}}\,
\frac{1}{\varepsilon+\omega-\varepsilon_{\mkq}+i\Sigma{''}_{\mkq}}
\nnnl
&=&\sum_{l=-\infty}^{\infty}\oint dz\,\frac{1}{z-\mu-ip_l}\,
\frac{1}{z- \varepsilon_{\nk}+i\Sigma{''}_{\nk}}\,
\frac{1}{z+\omega-\varepsilon_{\mkq}+i\Sigma{''}_{\mkq}}
\nnnl
&=&2\pi\,i\,\sum_{l=0}^{\infty}
\frac{1}{ip_l- (\varepsilon_{\nk}-\mu)+i\Sigma{''}_{\nk}}\,
\frac{1}{ip_l+\omega-(\varepsilon_{\mkq}-\mu)+i\Sigma{''}_{\mkq}}
\nnnl
&=&\beta\,\frac{-\psi\left[\tfrac{1}{2} - \tfrac{\beta}{2\pi i}\,(\varepsilon_{\nk}-\mu-i\Sigma{''}_{\nk})\right] + \psi\left[\tfrac{1}{2} - \tfrac{\beta}{2\pi i}\,(-\omega+\varepsilon_{\mkq}-\mu-i\Sigma{''}_{\mkq})\right]}{\varepsilon_\nk + \omega - \varepsilon_\mkq-i(\Sigma{''}_\nk-\Sigma{''}_\mkq)}\,.
\nnnl
\end{eqnarray}
\end{widetext}
With this result, we can easily evaluate the following integral from the second line of Eq.~\eqref{eq:energy_integral}:
\begin{widetext}
\begin{eqnarray}
\label{eq:energy_integral2}
&&\sum_{l=-\infty}^{\infty}\int d\varepsilon\,\frac{1}{\varepsilon-\mu-ip_l}\,
\frac{1}{\varepsilon- \varepsilon_{\nk}-i\Sigma{''}_{\nk}}\,
\frac{1}{\varepsilon+\omega-\varepsilon_{\mkq}-i\Sigma{''}_{\mkq}}\nnnl
&=&\left(\sum_{l=-\infty}^{\infty}\int d\varepsilon\,\frac{1}{\varepsilon-\mu+ip_l}\,
\frac{1}{\varepsilon- \varepsilon_{\nk}+i\Sigma{''}_{\nk}}\,
\frac{1}{\varepsilon+\omega-\varepsilon_{\mkq}+i\Sigma{''}_{\mkq}}\right)^*\nnnl
&=&\left(\sum_{l=-\infty}^{\infty}\int d\varepsilon\,\frac{1}{\varepsilon-\mu-ip_l}\,
\frac{1}{\varepsilon- \varepsilon_{\nk}+i\Sigma{''}_{\nk}}\,
\frac{1}{\varepsilon+\omega-\varepsilon_{\mkq}+i\Sigma{''}_{\mkq}}\right)^*\,,\nnnl
\end{eqnarray}
\end{widetext}
i.\,e.\,, the result of this integral is the complex conjugate of Eq.~\eqref{eq:energy_integral1}.

We can evaluate the following part of the second line of Eq.~\eqref{eq:energy_integral} by choosing an infinitely large upper semicircle as its contour:
\begin{widetext}
\begin{eqnarray}
\label{eq:energy_integral3}
&&-\sum_{l=-\infty}^{\infty}\int d\varepsilon\,\frac{1}{\varepsilon-\mu-ip_l}\,
\frac{1}{\varepsilon- \varepsilon_{\nk}+i\Sigma{''}_{\nk}}\,
\frac{1}{\varepsilon+\omega-\varepsilon_{\mkq}-i\Sigma{''}_{\mkq}}
\nnnl
&=&-\sum_{l=-\infty}^{\infty}\oint dz\,\frac{1}{z-\mu-ip_l}\,
\frac{1}{z- \varepsilon_{\nk}+i\Sigma{''}_{\nk}}\,
\frac{1}{z+\omega-\varepsilon_{\mkq}-i\Sigma{''}_{\mkq}}
\nnnl
&=&-2\pi\,i\,\sum_{l=0}^{\infty}
\frac{1}{ip_l- (\varepsilon_{\nk}-\mu)+i\Sigma{''}_{\nk}}\,
\frac{1}{ip_l+\omega-(\varepsilon_{\mkq}-\mu)-i\Sigma{''}_{\mkq}}\nnnl
&&-2\pi i\,\sum_{l=0}^{\infty}\left(
\frac{-1}{-\omega+\varepsilon_\mkq-\mu+i\Sigma{''}_\mkq-ip_l}
+\frac{-1}{-\omega+\varepsilon_\mkq-\mu+i\Sigma{''}_\mkq+ip_l}
\right)\nnnl
&&\quad\quad{ \times\,\frac{1}{\varepsilon_\nk+\omega-\varepsilon_\mkq-i(\Sigma{''}_\nk+\Sigma{''}_\mkq)}}\nnnl
&=&-2\pi\,i\,\sum_{l=0}^{\infty}
\frac{1}{ip_l- (\varepsilon_{\nk}-\mu)+i\Sigma{''}_{\nk}}\,
\frac{1}{ip_l+\omega-(\varepsilon_{\mkq}-\mu)-i\Sigma{''}_{\mkq}}\nnnl
&&-2\pi i\,\sum_{l=0}^{\infty}
\frac{1}{ip_l+\omega-\varepsilon_\mkq+\mu-i\Sigma{''}_\mkq}
\,\frac{1}{ip_l-\omega+\varepsilon_\mkq-\mu+i\Sigma{''}_\mkq}\nnnl
&&\quad\quad{ \times\,\frac{2(-\omega+\varepsilon_\mkq-\mu+i\Sigma{''}_\mkq)}{\varepsilon_\nk+\omega-\varepsilon_\mkq-i(\Sigma{''}_\nk+\Sigma{''}_\mkq)}}\nnnl
&=&\beta\,\frac{\psi\left[\tfrac{1}{2} - \tfrac{\beta}{2\pi i}\,(\varepsilon_{\nk}-\mu-i\Sigma{''}_{\nk})\right] - \psi\left[\tfrac{1}{2} - \tfrac{\beta}{2\pi i}\,(-\omega+\varepsilon_{\mkq}-\mu+i\Sigma{''}_{\mkq})\right]}{\varepsilon_\nk + \omega - \varepsilon_\mkq-i(\Sigma{''}_\nk+\Sigma{''}_\mkq)}\nnnl
&&+\beta\,\frac{\psi\left[\tfrac{1}{2} - \tfrac{\beta}{2\pi i}\,(-\omega+\varepsilon_{\mkq}-\mu+i\Sigma{''}_{\mkq})\right] - \psi\left[\tfrac{1}{2} + \tfrac{\beta}{2\pi i}\,(-\omega+\varepsilon_{\mkq}-\mu+i\Sigma{''}_{\mkq})\right]}{\varepsilon_\nk + \omega - \varepsilon_\mkq-i(\Sigma{''}_\nk+\Sigma{''}_\mkq)}\nnnl
&=&\beta\,\frac{\psi\left[\tfrac{1}{2} - \tfrac{\beta}{2\pi i}\,(\varepsilon_{\nk}-\mu-i\Sigma{''}_{\nk})\right]  - \psi\left[\tfrac{1}{2} + \tfrac{\beta}{2\pi i}\,(-\omega+\varepsilon_{\mkq}-\mu+i\Sigma{''}_{\mkq})\right]}{\varepsilon_\nk + \omega - \varepsilon_\mkq-i(\Sigma{''}_\nk+\Sigma{''}_\mkq)}\,,\nnnl
\end{eqnarray}
\end{widetext}
where the first and second lines on the right-hand side of the third equality were obtained from the residues of the poles of $1/(z-\mu-ip_l)$ and $1/(z+\omega-\varepsilon_\mkq-i\Sigma{''}_\mkq)$, respectively.
With this result, we can easily evaluate the following integral in the second line of Eq.~\eqref{eq:energy_integral}:
\begin{widetext}
\begin{eqnarray}
\label{eq:energy_integral4}
&&-\sum_{l=-\infty}^{\infty}\int d\varepsilon\,\frac{1}{\varepsilon-\mu-ip_l}\,
\frac{1}{\varepsilon- \varepsilon_{\nk}-i\Sigma{''}_{\nk}}\,
\frac{1}{\varepsilon+\omega-\varepsilon_{\mkq}+i\Sigma{''}_{\mkq}}
\nnnl
&=&\left(-\sum_{l=-\infty}^{\infty}\int d\varepsilon\,\frac{1}{\varepsilon-\mu+ip_l}\,
\frac{1}{\varepsilon- \varepsilon_{\nk}+i\Sigma{''}_{\nk}}\,
\frac{1}{\varepsilon+\omega-\varepsilon_{\mkq}-i\Sigma{''}_{\mkq}}\right)^*
\nnnl
&=&\left(-\sum_{l=-\infty}^{\infty}\int d\varepsilon\,\frac{1}{\varepsilon-\mu-ip_l}\,
\frac{1}{\varepsilon- \varepsilon_{\nk}+i\Sigma{''}_{\nk}}\,
\frac{1}{\varepsilon+\omega-\varepsilon_{\mkq}-i\Sigma{''}_{\mkq}}\right)^*\,,
\nnnl
\end{eqnarray}
\end{widetext}
i.\,e.\,, the result of this integral is the complex conjugate of Eq.~\eqref{eq:energy_integral3}.

Plugging the results in Eqs.~\eqref{eq:energy_integral1}-\eqref{eq:energy_integral4} into Eq.~\eqref{eq:energy_integral}, we obtain
\begin{widetext}
\begin{eqnarray}
\label{eq:energy_integral_result}
&&\sum_{l=-\infty}^{\infty}\int d\varepsilon\,\frac{1}{\varepsilon-\mu-ip_l}\,\frac{\Sigma{''}_{\nk}}{(\varepsilon- \varepsilon_{\nk})^2+\Sigma{''}_{\nk}^2}\,
\frac{\Sigma{''}_{\mkq}}{(\varepsilon+\omega-\varepsilon_{\mkq})^2+\Sigma{''}_{\mkq}^2}
\nnnl
&&=-\frac{\beta}{2}\,{\rm Re\,}\left\{ \frac{-\psi\left[\tfrac{1}{2} - \tfrac{\beta}{2\pi i}\,(\varepsilon_{\nk}-\mu-i\Sigma{''}_{\nk})\right] + \psi\left[\tfrac{1}{2} - \tfrac{\beta}{2\pi i}\,(-\omega+\varepsilon_{\mkq}-\mu-i\Sigma{''}_{\mkq})\right]}{\varepsilon_\nk + \omega - \varepsilon_\mkq-i(\Sigma{''}_\nk-\Sigma{''}_\mkq)} \right.
\nnnl
&&+\left. \frac{\psi\left[\tfrac{1}{2} - \tfrac{\beta}{2\pi i}\,(\varepsilon_{\nk}-\mu-i\Sigma{''}_{\nk})\right]  - \psi\left[\tfrac{1}{2} + \tfrac{\beta}{2\pi i}\,(-\omega+\varepsilon_{\mkq}-\mu+i\Sigma{''}_{\mkq})\right]}{\varepsilon_\nk + \omega - \varepsilon_\mkq-i(\Sigma{''}_\nk+\Sigma{''}_\mkq)}
\right\}
\nnnl
\end{eqnarray}
\end{widetext}

Now, plugging Eq.~\eqref{eq:energy_integral_result} back into Eq.~\eqref{eq:Pi_omega2}, we finally obtain Eq.~\eqref{eq:Pi_omega3}.

\subsection{Limiting cases of Eq.~\eqref{eq:Pi_omega3}}
\label{sec:limiting_cases}
In this section, we derive Eqs.~\eqref{eq:Pi_omega_limit1_short} and~\eqref{eq:Pi_omega_limit2_short}.
To discuss the limiting cases of Eq.~\eqref{eq:Pi_omega3}, we first show from Eq.~\eqref{eq:digamma} that for real $\varepsilon$
\begin{align}
\label{eq:Im_digamma}
&{\rm Im\,}\,\psi\left[\tfrac{1}{2} - \tfrac{\beta}{2\pi i}\,(\varepsilon-\mu)\right]
\nnnl
&=-{\rm Im\,}\,\sum_{l=0}^\infty \frac{1}{l+\frac{1}{2}- \tfrac{\beta}{2\pi i}\,(\varepsilon-\mu)}
\nnnl
&=\frac{i}{2}\,\sum_{l=0}^\infty \left[
\frac{1}{l+\frac{1}{2}- \tfrac{\beta}{2\pi i}\,(\varepsilon-\mu)}
-\frac{1}{l+\frac{1}{2}+ \tfrac{\beta}{2\pi i}\,(\varepsilon-\mu)}
\right]
\nnnl
&=\frac{i}{2}\,\sum_{l=-\infty}^\infty
\frac{1}{l+\frac{1}{2}- \tfrac{\beta}{2\pi i}\,(\varepsilon-\mu)}
\nnnl
&=\frac{\pi}{\beta}\,\sum_{l=-\infty}^\infty \frac{1}{\varepsilon-\mu-ip_l}
\nnnl
&=\pi\left[\frac{1}{2}-f(\varepsilon)\right]\,,
\end{align}
where in the last equality, we have used Eq.~\eqref{eq:f_Matsubara}.

Let us find the $\Sigma{''}_\nk\to0^+$ limit of Eq.~\eqref{eq:Pi_omega3}.
\begin{widetext}
\begin{eqnarray}
\label{eq:Pi_omega_limit1}
&&{\rm Im\,} \Pi^{\rm NA}_{\qnu}(\omega+i0^+)=-\frac{1}{2\pi}\,\sum_{mn}\,\int \frac{d{\bf k}}{\Omega_{\rm BZ}} 
|g_{mn\nu}({\bf k},{\bf q})|^2\,\times\nnnl
&&{\rm Re\,}\left\{ \frac{-\psi\left[\tfrac{1}{2} - \tfrac{\beta}{2\pi i}\,(\varepsilon_{\nk}-\mu)\right] + \psi\left[\tfrac{1}{2} - \tfrac{\beta}{2\pi i}\,(-\omega+\varepsilon_{\mkq}-\mu-i\Sigma{''}_{\mkq})\right]}{\varepsilon_\nk + \omega - \varepsilon_\mkq+i\Sigma{''}_\mkq} \right.
\nnnl
&&+\frac{\psi\left[\tfrac{1}{2} - \tfrac{\beta}{2\pi i}\,(\varepsilon_{\nk}-\mu)\right]  - \psi\left[\tfrac{1}{2} + \tfrac{\beta}{2\pi i}\,(-\omega+\varepsilon_{\mkq}-\mu+i\Sigma{''}_{\mkq})\right]}{\varepsilon_\nk + \omega - \varepsilon_\mkq-i\Sigma{''}_\mkq}
\nnnl
&&+\frac{\psi\left[\tfrac{1}{2} - \tfrac{\beta}{2\pi i}\,(\varepsilon_{\nk}+\omega-\mu)\right]-\psi\left[\tfrac{1}{2} - \tfrac{\beta}{2\pi i}\,(\varepsilon_{\mkq}-\mu-i\Sigma{''}_{\mkq})\right] }{\varepsilon_\nk + \omega - \varepsilon_\mkq+i\Sigma{''}_\mkq}
\nnnl
&&+\left.\frac{- \psi\left[\tfrac{1}{2} + \tfrac{\beta}{2\pi i}\,(\varepsilon_{\nk}+\omega-\mu)\right]+\psi\left[\tfrac{1}{2} - \tfrac{\beta}{2\pi i}\,(\varepsilon_{\mkq}-\mu-i\Sigma{''}_{\mkq})\right] }{\varepsilon_\nk + \omega - \varepsilon_\mkq+i\Sigma{''}_\mkq}\right\}
\nnnl
&&=-\frac{1}{2\pi}\,\sum_{mn}\,\int \frac{d{\bf k}}{\Omega_{\rm BZ}} 
|g_{mn\nu}({\bf k},{\bf q})|^2\,\times\nnnl
&&{\rm Re\,}\left\{-\psi\left[\tfrac{1}{2} - \tfrac{\beta}{2\pi i}\,(\varepsilon_{\nk}-\mu)\right]\,2i\,
{\rm Im\,} \frac{1}{\varepsilon_\nk + \omega - \varepsilon_\mkq+i\Sigma{''}_\mkq}
\right.
\nnnl
&&+\left.2i\,{\rm Im\,}\psi\left[\tfrac{1}{2} - \tfrac{\beta}{2\pi i}\,(\varepsilon_{\nk}+\omega-\mu)\right] \frac{1}{\varepsilon_\nk + \omega - \varepsilon_\mkq+i\Sigma{''}_\mkq}\right\}
\nnnl
&&=-\frac{1}{\pi}\,\sum_{mn}\,\int \frac{d{\bf k}}{\Omega_{\rm BZ}} 
|g_{mn\nu}({\bf k},{\bf q})|^2\,\times\nnnl
&&{\rm Im\,}\left\{\psi\left[\tfrac{1}{2} - \tfrac{\beta}{2\pi i}\,(\varepsilon_{\nk}-\mu)\right]
-\psi\left[\tfrac{1}{2} - \tfrac{\beta}{2\pi i}\,(\varepsilon_{\nk}+\omega-\mu)\right]\right\}\nnnl
&&{\times\,{\rm Im\,} \frac{1}{\varepsilon_\nk + \omega - \varepsilon_\mkq+i\Sigma{''}_\mkq}}
\nnnl
&&=-\pi\,\sum_{mn}\,\int \frac{d{\bf k}}{\Omega_{\rm BZ}} 
|g_{mn\nu}({\bf k},{\bf q})|^2\,\left[f(\varepsilon_\nk)-f(\varepsilon_\nk+\omega)\right]\,
\delta_{\mkq} (\varepsilon_\nk+\omega)\,,
\end{eqnarray}
\end{widetext}
where in the last equality we have used Eqs.~\eqref{eq:delta_mk} and~\eqref{eq:Im_digamma}.

Now, similarly, let us find the $\Sigma{''}_\mkq\to0^+$ limit of Eq.~\eqref{eq:Pi_omega3}.
\begin{widetext}
\begin{eqnarray}
\label{eq:Pi_omega_limit2}
&&{\rm Im\,} \Pi^{\rm NA}_{\qnu}(\omega+i0^+)=-\frac{1}{2\pi}\,\sum_{mn}\,\int \frac{d{\bf k}}{\Omega_{\rm BZ}} 
|g_{mn\nu}({\bf k},{\bf q})|^2\,\times\nnnl
&&{\rm Re\,}\left\{ \frac{-\psi\left[\tfrac{1}{2} - \tfrac{\beta}{2\pi i}\,(\varepsilon_{\nk}-\mu-i\Sigma{''}_{\nk})\right] + \psi\left[\tfrac{1}{2} - \tfrac{\beta}{2\pi i}\,(-\omega+\varepsilon_{\mkq}-\mu)\right]}{\varepsilon_\nk + \omega - \varepsilon_\mkq-i\Sigma{''}_\nk} \right.
\nnnl
&&+\frac{\psi\left[\tfrac{1}{2} - \tfrac{\beta}{2\pi i}\,(\varepsilon_{\nk}-\mu-i\Sigma{''}_{\nk})\right]  - \psi\left[\tfrac{1}{2} + \tfrac{\beta}{2\pi i}\,(-\omega+\varepsilon_{\mkq}-\mu)\right]}{\varepsilon_\nk + \omega - \varepsilon_\mkq-i\Sigma{''}_\nk}
\nnnl
&&+\frac{ \psi\left[\tfrac{1}{2} - \tfrac{\beta}{2\pi i}\,(\varepsilon_{\nk}+\omega-\mu-i\Sigma{''}_{\nk})\right]-\psi\left[\tfrac{1}{2} - \tfrac{\beta}{2\pi i}\,(\varepsilon_{\mkq}-\mu)\right]}{\varepsilon_\nk + \omega - \varepsilon_\mkq-i\Sigma{''}_\nk}
\nnnl
&&+\left.\frac{- \psi\left[\tfrac{1}{2} + \tfrac{\beta}{2\pi i}\,(\varepsilon_{\nk}+\omega-\mu+i\Sigma{''}_{\nk})\right]+\psi\left[\tfrac{1}{2} - \tfrac{\beta}{2\pi i}\,(\varepsilon_{\mkq}-\mu)\right] }{\varepsilon_\nk + \omega - \varepsilon_\mkq+i\Sigma{''}_\nk}\right\}
\nnnl
&&=-\frac{1}{2\pi}\,\sum_{mn}\,\int \frac{d{\bf k}}{\Omega_{\rm BZ}} 
|g_{mn\nu}({\bf k},{\bf q})|^2\,\times\nnnl
&&{\rm Re\,}\left\{
2i\,{\rm Im\,}\psi\left[\tfrac{1}{2} - \tfrac{\beta}{2\pi i}\,(-\omega+\varepsilon_{\mkq}-\mu)\right] \frac{1}{\varepsilon_\nk + \omega - \varepsilon_\mkq-i\Sigma{''}_\nk}
\right.
\nnnl
&&-\left.
\psi\left[\tfrac{1}{2} - \tfrac{\beta}{2\pi i}\,(\varepsilon_{\mkq}-\mu)\right]\,2i\,
{\rm Im\,} \frac{1}{\varepsilon_\nk + \omega - \varepsilon_\mkq-i\Sigma{''}_\nk}
\right\}
\nnnl
&&=-\frac{1}{\pi}\,\sum_{mn}\,\int \frac{d{\bf k}}{\Omega_{\rm BZ}} 
|g_{mn\nu}({\bf k},{\bf q})|^2\,\times\nnnl
&&{\rm Im\,}\left\{\psi\left[\tfrac{1}{2} - \tfrac{\beta}{2\pi i}\,(-\omega+\varepsilon_{\mkq}-\mu)\right]
-\psi\left[\tfrac{1}{2} - \tfrac{\beta}{2\pi i}\,(\varepsilon_{\mkq}-\mu)\right]\right\}\nnnl
&&{\times\,{\rm Im\,} \frac{1}{\varepsilon_\nk + \omega - \varepsilon_\mkq+i\Sigma{''}_\nk}}
\nnnl
&&=-\pi\,\sum_{mn}\,\int \frac{d{\bf k}}{\Omega_{\rm BZ}} 
|g_{mn\nu}({\bf k},{\bf q})|^2\,\left[f(-\omega+\varepsilon_\mkq)-f(\varepsilon_\mkq)\right]\,
\delta_{\nk} (-\omega+\varepsilon_\mkq)\,,
\end{eqnarray}
\end{widetext}
where in the last equality we have used Eqs.~\eqref{eq:delta_mk} and~\eqref{eq:Im_digamma}.

\subsection{Derivation of Eq.~\eqref{eq:Gamma_graphene}}
\label{sec:graphene_derivation}
{
Here, $\Omega_\textrm{BZ}=\frac{(2\pi)^2}{A}$, where $A=\frac{3\sqrt{3}}{2}b^2$ is the area of a unit cell. We are interested in the case $\textbf{q}=\textbf{k}'-\textbf{k}\to\textbf{0}$.
We consider the longitudinal-optical (LO) and transverse-optical (TO) $E_\textrm{2g}$ phonon modes. We first perform the angular integration in \textbf{k} space. Let us denote $\theta_\textbf{k}=\theta$. Importantly, the ${\rm Re\,}\left\{...\right\}$ part in Eq.~\eqref{eq:Pi_omega3} is independent of $\theta$ in the limit $\textbf{q}\to0$. Therefore, we can perform the two-dimensional integration over \textbf{k} first over $\theta$ for $|g_{mn\nu}(\textbf{k},\textbf{q})|^2$ and then over $k=\hbar \varepsilon/v_\textrm{F}$ for the rest of Eq.~\eqref{eq:Pi_omega3}. The analytic forms of the matrix elements are presented in Tab.~I of Ref.~21.

The angular integration results in
\begin{align} \label{eq:g2_ang_int}
&\frac{1}{2\pi}\int_0^{2\pi} d\theta\, |g_{mn\nu}(\textbf{k},\textbf{q})|^2\nnnl
=&
\left(\frac{3}{2}\,\eta\,\sqrt{\frac{\hbar}{2M_{\rm C}\omega_\Gamma^{{\rm E}_{2g}}}}\right)^2\cdot\frac{1}{\pi}\int_0^{2\pi} d\theta\,\sin^2(\theta-\theta_\textbf{q}+\theta_{mn\nu})\nnnl
=&\left(\frac{3}{2}\,\eta\,\sqrt{\frac{\hbar}{2M_{\rm C}\omega_\Gamma^{{\rm E}_{2g}}}}\right)^2=\bar{g}^2\,,
\end{align}
where $\theta_{mn\nu}$ is a constant which depends on the electron bands and phonon branch.

Now, from Eqs.~(11) and~(16) and the replacement $\varepsilon=\hbar v_\textrm{F}k$, the two-dimensional integration in momentum space, together with the consideration of spin and valley degeneracy of $4$, results in
\begin{align} \label{eq:g2_radial}
\Gamma_\qnu&=\frac{1}{\pi}\sum_{mn}\int \frac{d\textbf{k}}{\Omega_\textrm{BZ}}|g_{mn\nu}(\textbf{k},\textbf{q})|^2\{...\}\nnnl
&=\frac{1}{\pi}\cdot\frac{3\sqrt{3}b^2}{2(2\pi)^2}\cdot4\cdot2\pi\,\bar{g}^2\sum_{m,\,n=\,\{-1,+1\}}\,\int k\, dk\,\{...\}\nnnl
&=\frac{3\sqrt{3}}{\pi^2}\cdot\left(\frac{b\bar{g}}{\hbar v_\textrm{F}}\right)^2\sum_{m,\,n=\,\{-1,+1\}}\,\int \varepsilon\, d\varepsilon\,\{...\}\,,
\end{align}
which completes the proof of Eq.~\eqref{eq:Gamma_graphene}.
}

\bibliography{main}

\end{document}